\documentclass[aps,prl,twocolumn,showpacs,superscriptaddress,floatfix]{revtex4-1}
\usepackage{caption}
\usepackage{graphicx} 
\usepackage{subfigure}  
\usepackage{dcolumn}   
\usepackage{bm}  
\usepackage{epstopdf} 
\usepackage{amsmath}      
\usepackage{amssymb}   

\begin{document}

\title{Electronic structure of tungsten-doped vanadium dioxide}
\author{Jamie ~M.~Booth}
\email{jamie.booth@rmit.edu.au}

\affiliation{Australian Research Council Centre of Excellence for Exciton Science, School of Science, RMIT University, Melbourne 3001, VIC, Australia}

\author{Daniel ~W.~Drumm}
\affiliation{Australian Research Council Centre of Excellence for Nanoscale BioPhotonics, School of Science, RMIT University, Melbourne 3001, VIC, Australia}

\author{Phil ~S.~Casey}
\affiliation{CSIRO Manufacturing, Clayton VIC 3168, Australia}

\author{Jackson ~S.~Smith}
\affiliation{Theoretical Chemical and Quantum Physics, School of Science, RMIT University, Melbourne VIC 3001, Australia}

\author{Salvy ~P.~Russo}
\affiliation{Australian Research Council Centre of Excellence for Exciton Science, School of Science, RMIT University, Melbourne 3001, VIC, Australia}

\date{\today}

\begin{abstract}
A common method of adjusting the metal-insulator transition temperature of M$_{1}$ VO$_{2}$ is via disruption of the Peierls pairing by doping, or inputting stress or strain. However, since adding even small amounts of dopants will change the band structure, it is unclear how doped VO$_{2}$ retains its insulating character observed in experiments. While strong correlations may be responsible for maintaining a gap, theoretical evidence for this has been very difficult to obtain due to the complexity of the many-body problem involved. In this work we use GW calculations modified to include strong local \textbf{k}-space interactions to investigate the changes in band structure from tungsten doping. We find that the combination of carrier doping and the experimentally observed structural defects introduced by inclusion of tungsten are consistent with a change from band-like to Mott-insulating behavior.
\end{abstract}

\pacs{71.30.+h,71.27.+a,74.20.Pq,75.10.-b,71.20.-b}
\maketitle

\textit{\textbf{Introduction}}: Vanadium dioxide is considered a prototypical strongly correlated material, which undergoes a first order insulator metal transition at $\sim$ 340 K from a low temperature monoclinic structure to a high temperature, metallic tetragonal structure \cite{Goodenough1971}. In the last decade, advances in nanofabrication techniques have seen an explosion of interest in devices based on VO$_{2}$ nanobeams \cite{Wei2012,Park2013}, which show enormous potential for sensors \cite{Guo2011} and even new transistor gates \cite{Nakano2012}. Dynamical Mean Field Theory \cite{Biermann2005} and GW calculations \cite{Gatti2007} suggest that pure VO$_{2}$'s insulating character results from the three-dimensional pairings that occur in the transition from tetragonal to monoclinic which fill the valence band, with strong correlations in the metallic form responsible for driving the transition to the insulating state \cite{Biermann2005}. 

Many methods of controlling the critical temperature revolve around disrupting this pairing, such as the input of stress or strain \cite{Wei2009}, or doping \cite{Lawton1995}. Doping with tungsten has long been known to reduce T$_{c}$ by $\sim$ 23 K per atomic percent of tungsten \cite{Lawton1995}. However, when viewed through the lens of band theory, this presents something of a paradox, as doping carriers into pure VO$_{2}$ should result in a metallic structure. Experiments confirm however, that VO$_{2}$ doped with $<$ 10 \% tungsten remains insulating \cite{Tang1985}. In addition, photoemission experiments reveal that tungsten assumes the place of a vanadium atom in the VO$_{2}$ lattice, but does \textit{not} Peierls pair with neighbouring atoms \cite{Tang1985}, despite having the valence electrons to do so. Rather, the local environment of the tungsten is tetragonal, while the structure retains an overall $P2_{1}/c$ symmetry. This is in contrast to doping with chromium, which despite being in the same group as tungsten, promotes the formation of the Mott insulating M$_{2}$ structure at dopant percentages as low as 0.5 \% \cite{Marezio1971}. 

\begin{figure}[h!]
  \includegraphics[width=0.6\columnwidth]{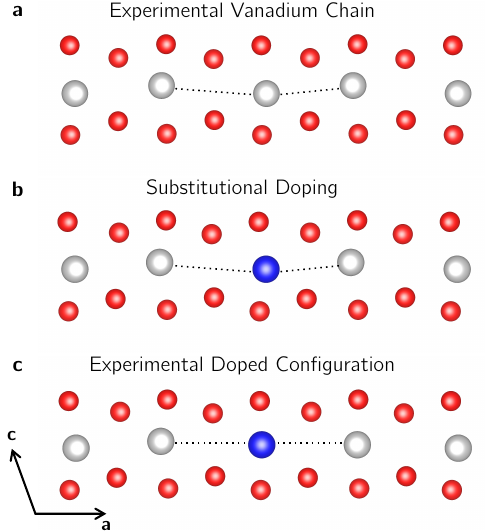}
  \caption{\raggedright{Color Online a) Undoped vanadium chain running along the monoclinic a-axis, b) substitutionally doped chain used in the calculations of Figures 2 and 3, and c) experimentally determined doped geometry used in the calculations of Figure 4.}}
\label{fig:Fig1}
\end{figure}

Figure \ref{fig:Fig1} compares the different geometries involved: Figure \ref{fig:Fig1}a illustrates the Peierls paired and antiferroelectrically distorted vanadium chain in the undoped compound. Figure \ref{fig:Fig1}b illustrates a substitutionally doped chain, in which the tungsten adopts the position of a vanadium atom.  Figure \ref{fig:Fig1}c corresponds to the experimentally observed \cite{Tang1985,Booth2009a} tungsten environment; the tungsten atom sits equidistant from the neighbouring vanadium atoms, and in the center of the oxygen octadehron, i.e. with no Peierls pairing or antiferroelectric distortion. Given that the insulating state occurs via the vanadium atoms in the tetragonal structure forming pairs and undergoing an antiferroelectric distortion \cite{Goodenough1971}, disruption of this structural rearrangement as per Figure 1c, combined with the extra carriers might appear from a purely band theoretical view to transform tungsten-doped VO$_{2}$ to a metallic structure. Unless of course the experimentally observed insulating behavior in the doped structure arises due to strong correlations in the partially filled band.
\begin{figure}[h!]
  \subfigure{\includegraphics[width=0.95\columnwidth]{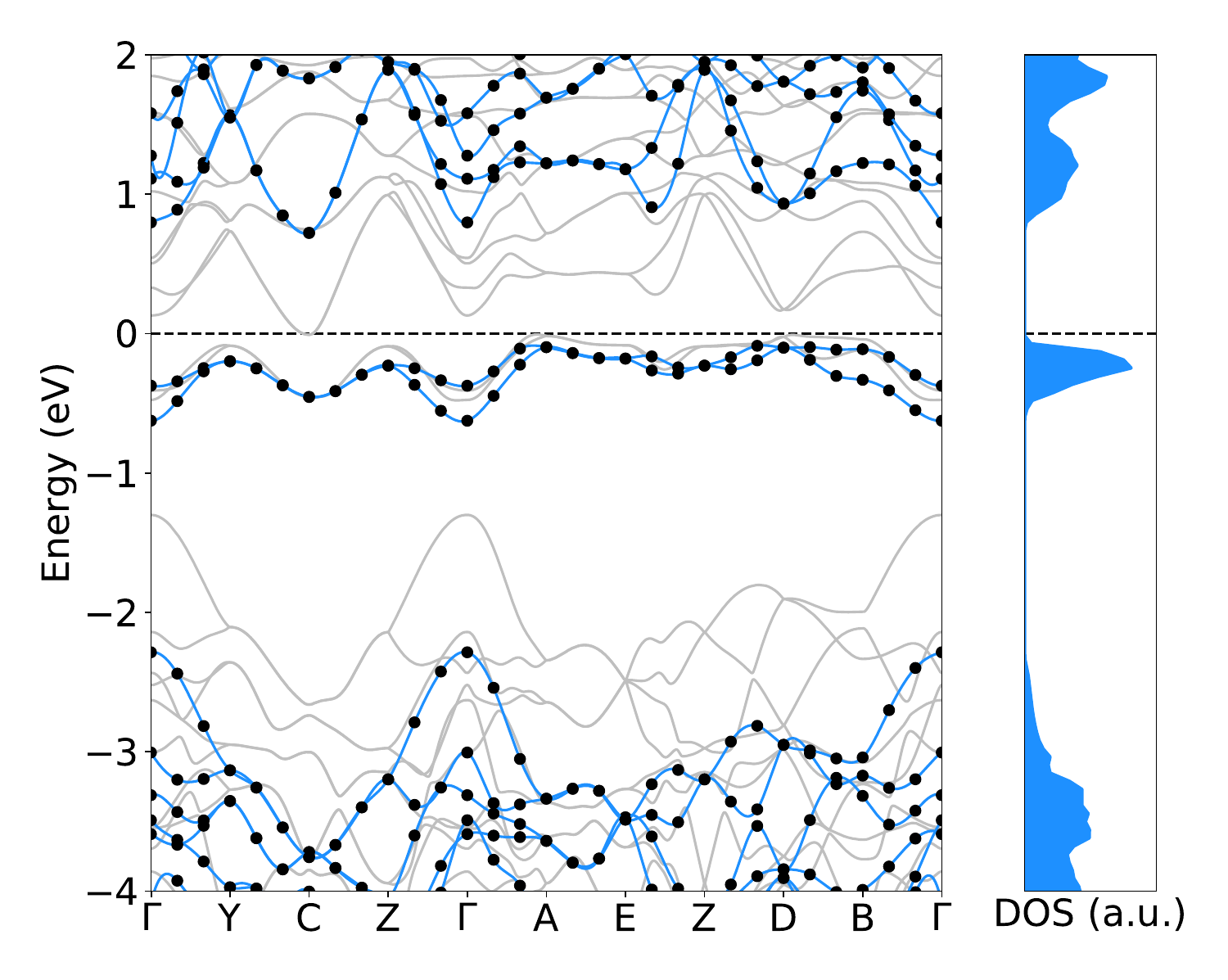}}\textbf{{\textsf{a}}}
  \subfigure{\includegraphics[width=0.95\columnwidth]{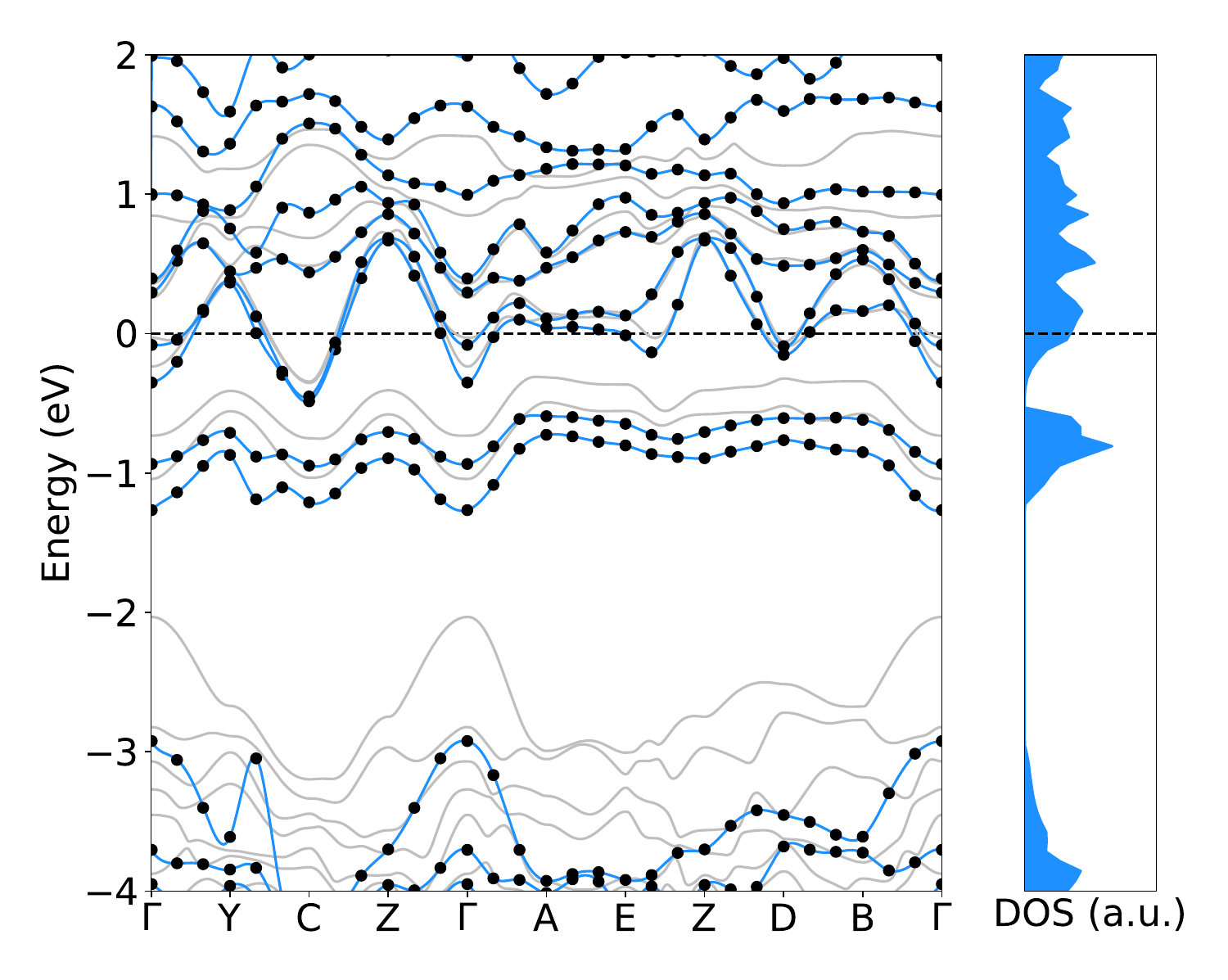}}\textbf{{\textsf{b}}}
  \subfigure{\includegraphics[width=0.8\columnwidth]{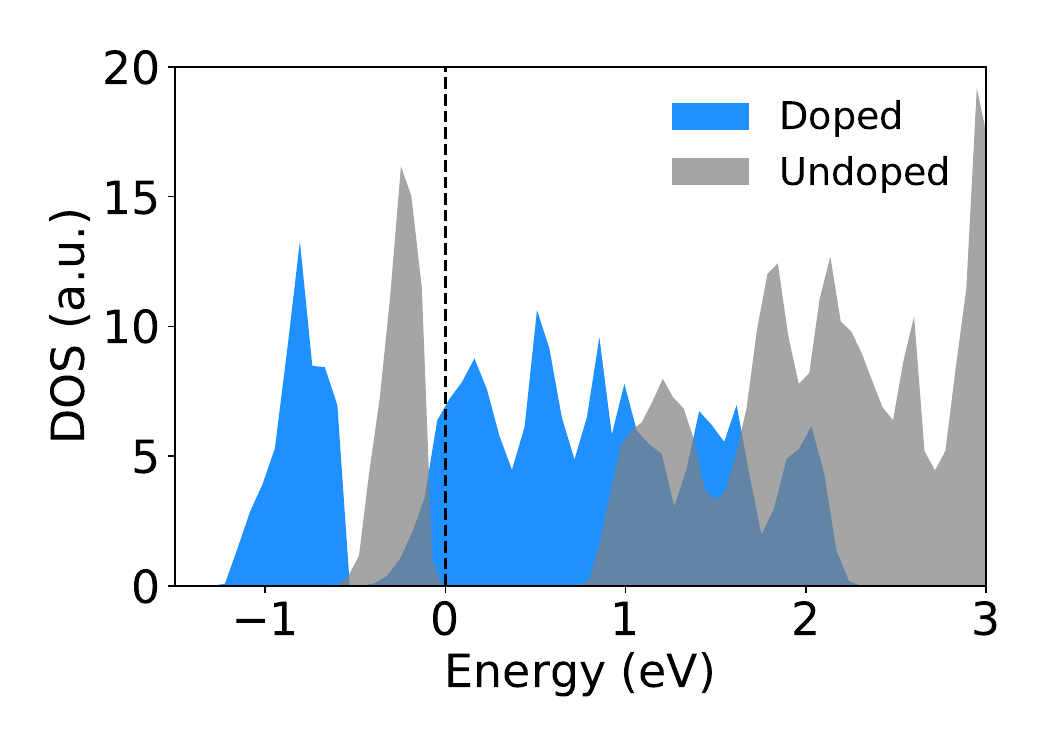}}\textbf{{\textsf{c}}}
  \caption{\raggedright{Color Online a) Left panel: G$_{0}$W$_{0}$ band structure of M$_{1}$ VO$_{2}$ (black circles fitted with blue splines) and corresponding DFT bands (gray lines), Right panel: corresponding G$_{0}$W$_{0}$ density of states (DOS)b) Left panel: G$_{0}$W$_{0}$ (black circles fitted with blue splines) and DFT (gray lines) band structures of tungsten doped VO$_{2}$, Right panel: corresponding G$_{0}$W$_{0}$ DOS, and c) comparison of the undoped and substitutionally doped G$_{0}$W$_{0}$ densities of states.}}
\label{fig:Fig2}
\end{figure}

The resolution of such a paradox from a computational perspective has to-date been out of reach, due to the failure of \textit{ab initio} methods to reproduce Mott physics \cite{Burke2012}. Band theory in particular fails spectacularly \cite{Imada1998}, such that even qualitative predictions are almost always inaccurate. For example, DFT characteristically predicts Mott insulators to be metallic \cite{Anisimov1991,Burke2012}, a significant issue when designing materials to take advantage of their insulating behavior. Recently however, it was revealed that GW calculations can be modified to include strong local \textbf{k}-space correlations (the ``Partially Screened GW" method or PS-GW) \cite{Booth2016a}, and can reproduce two significant characteristics of Mott physics: the splitting of partially filled bands into the upper and lower Hubbard bands, and the giant transfer of spectral weight upon carrier doping. The ability of this approach to reproduce these characteristics means that it is now possible to investigate systems in which the rigid band model fails, and in this work we apply it to the study of tungsten-doped VO$_{2}$.

\textit{\textbf{Methods}}: The M$_{1}$ VO$_{2}$ structure \cite{Andersson1954} was first relaxed to the ground state using PBE-GGA \cite{Perdew1996} Density Functional Theory (DFT) and the Brillouin zone integration method of Bloechl \textit{et al.} \cite{Bloechl1994}. $6\times6\times6$, $4\times6\times6$, $4\times4\times4$ Monkhorst-Pack \cite{Monkhorst1976} \textbf{k}-space grids were used for the 25 \%, 12.5 \% and 3.1 \% tungsten-doped structures respectively. These consisted of one, two and eight VO$_{2}$ unit cells in $1\times1\times1$, $2\times1\times1$, and $2\times2\times2$ configurations in which a single vanadium atom was replaced with tungsten. A discussion motivating the use of geometry relaxation to generate input structures in presented in the Supporting Information. The GW calculations were performed using the implementation of Shishkin and Kresse \cite{Shishkin2006,Shishkin2007} as contained in the Vienna Ab Initio Simulation Package (VASP) \cite{Kresse1996}, after first calculating input wavefunctions using DFT with the PBE\cite{Perdew1996} functional. The standard, or unmodified, GW calculations were performed on a frequency grid of 30 points, using a cutoff energy of 200 eV. Strong correlations were included by setting the derivatives of the wavefunctions with respect to the \textbf{k}-point grid to zero (the PS-GW method, see Booth \textit{et al.} \cite{Booth2016a}). Five self-consistency steps were used for the 12.5 \% doped structures, while four and two steps were used for the 12.5 \% and 3.1 \% doped structures respectively. All strongly correlated calculations were performed at a single frequency point, $\omega$=0 \cite{{Booth2016a}}.

\textit{\textbf{Discussion}}: Figure \ref{fig:Fig2}a illustrates the results of unmodified G$_{0}$W$_{0}$ calculations (i.e. not PS-GW) on the M$_{1}$ structure of VO$_{2}$ (corresponding to the vanadium chain environment of Figure \ref{fig:Fig1}a). In all plots presented in this work the Fermi level is set to zero energy. The GW band structure (black filled circles) exhibits substantial splitting of the valence and conduction bands at E$_{F}$ in contrast to the DFT bands (gray lines), in which the gap manifests as a slight splitting of the bands at E$_{F}$. The calculated GW gap is $\sim$ 0.70 eV, in excellent agreement with photoemission measurements (also $\sim$ 0.70 eV \cite{Shin1990}). Therefore, for M$_{1}$ VO$_{2}$, screening the Hartree-Fock interaction with the dielectric matrix resulting from non-interacting Green functions is sufficient to reproduce the experimental characteristics of the undoped structure. 

\begin{figure}[th!]
  \subfigure{\includegraphics[width=0.95\columnwidth]{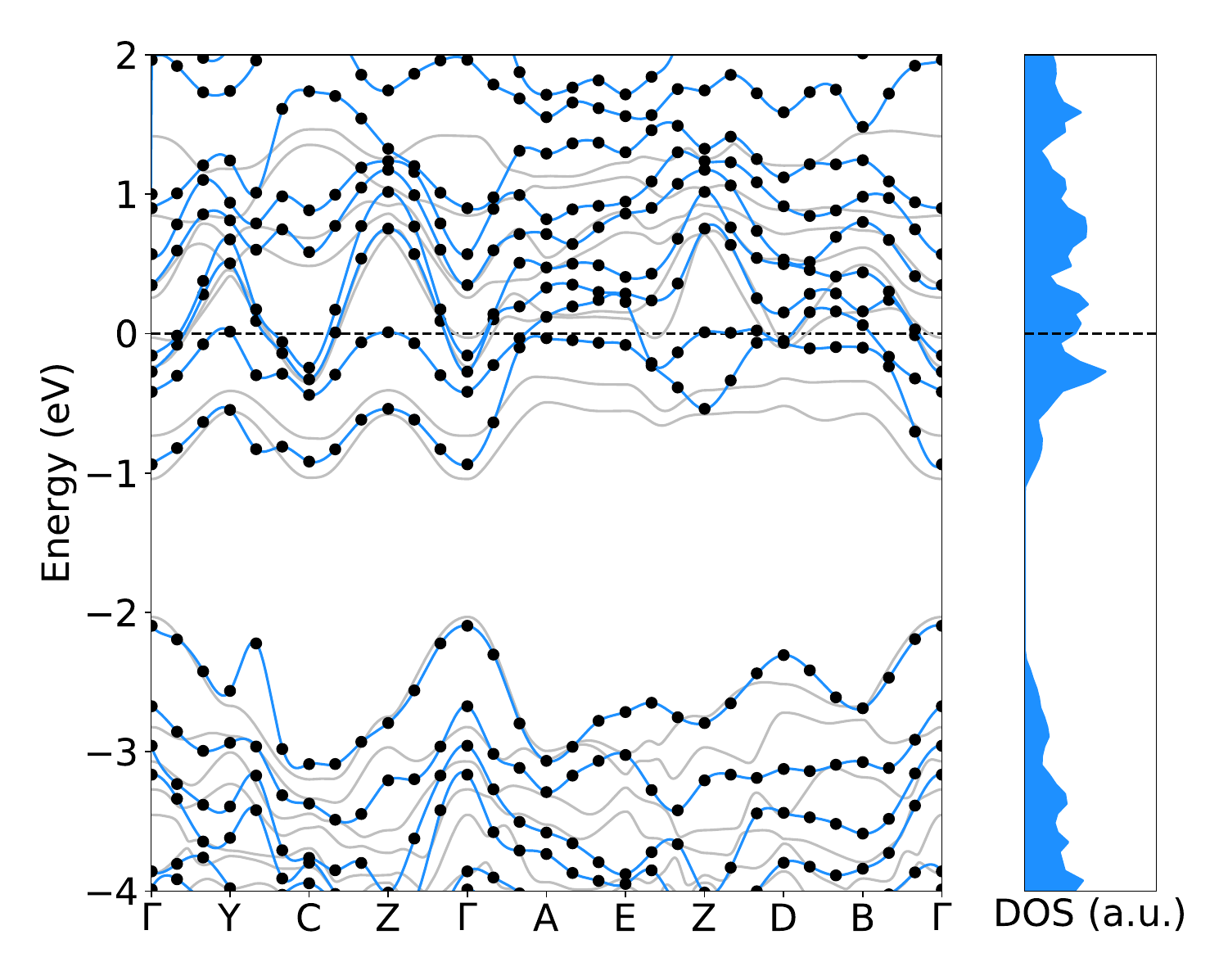}}\textbf{\textsf{a}}
  \subfigure{\includegraphics[width=0.8\columnwidth]{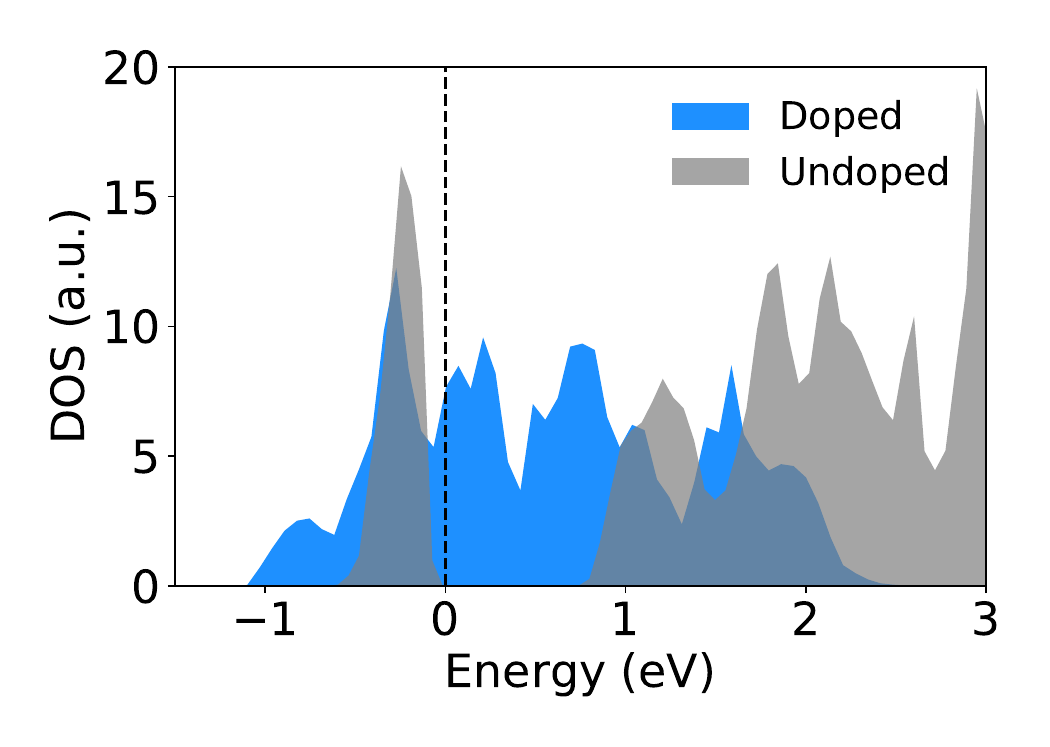}}\textbf{{\textsf{b}}}
  \caption{\raggedright{Color Online a) Left Panel: G$_{0}$W$_{0}$ (black circles fitted with blue splines) band structure of the tungsten doping configuration of Figure 1c and the corresponding DFT bands (gray lines), Right Panel: correponding G$_{0}$W$_{0}$ density of states, b) comparison of the undoped (Figure 1a) and doped (Figure 1c) G$_{0}$W$_{0}$ densities of states.}}
\label{fig:Fig3}
\end{figure}

However, Figure \ref{fig:Fig2}b indicates that for substitutionally tungsten-doped VO$_{2}$ this is not the case.  This figure plots an unmodified GW calculation of a doped configuration identical to that of Figure 1b and it is clear that the Fermi level now intersects the conduction band minima at $\Gamma$, $C$, $E$ and $D$. The curvatures of the DFT bands are in reasonable agreement with the splines fitted to the GW data which indicates that for a doped configuration of this kind there will be a continuum of states available above E$_{F}$, as the bands vary smoothly in \textbf{k}-space. Therefore, since pure VO$_{2}$ has a filled valence band, if we assume band theory to be valid in the doped case, a dopant level on the order of a few atomic percent tungsten will result in a metallic structure due to the large number of states above E$_{F}$.

This situation embodies the failure of the rigid band model for strongly correlated materials. While pure VO$_{2}$ has a filled valence band (Figure \ref{fig:Fig2}a) any doped states will necessarily inhabit the conduction band. However the conduction states consist of a broad mix of bands, corresponding to a peak in the DOS of approximately 2.5 eV. There are no gaps, and therefore shifting the Fermi level into this peak will create a metallic structure. However, as stated above, the experimentally observed doped configuration is that of Figure 1c \cite{Tang1985}, in which the tungsten atom adopts a local environment which is tetragonal, not the Peierls-paired monoclinic structure of the pure form of M$_{1}$ VO$_{2}$.

Thus, an investigation of the band structure of the linearized, experimentally observed geometry may shed some light on the nature of this insulating doped phase. Figure \ref{fig:Fig3}a illustrates the band structure calculated using the G$_{0}$W$_{0}$ method of the doped configuration of Figure \ref{fig:Fig1}c. A comparison with Figures \ref{fig:Fig3}a and \ref{fig:Fig2}b reveals that rather than this structure  appearing insulating, the splitting between the original valence and conduction bands of the undoped VO$_{2}$ structure has vanished completely. The two filled valence $d$-bands which in both the undoped form, and the subsitutionally doped form are separated from the conduction bands now overlap significantly with them. Figure \ref{fig:Fig3}b contrasts the densities of states of the undoped and experimentally doped form, and it is clear that the original band gap has completely closed, with a broad peak in the density of states forming which straddles the Fermi level.

This indicates an apparent paradox. If both the undoped and doped structures are experimentally found to be insulating, and standard GW calculations predict a band gap in the undoped form in good agreement with experiment, then the vanishing of any splitting between the bands in the doped form would suggest a transition to a highly metallic state. While the amount of dopant in the structure whose bands are presented in Figure 3a is quite high: 25 \%, from a band theory perspective the complete disappearance of the band splitting in the original insulating structure seems totally incompatible with an insulating structure. In addition, the shift of the Fermi level due to the extra valence electron into the broad conduction band suggests that even if any splitting were maintained, such as in the substitutionally doped structure (see Figure \ref{fig:Fig2}b), that it would not result in a band insulator.

The reduction in splitting of the valence bands is however compatible with the atomic arrangement of the tetragonal structure. Tetragonal VO$_{2}$ does not exhibit Peierls pairing of the vanadium atoms, unlike the structures of Figures 1a-b. Instead the vanadium atoms are evenly spaced along the tetragonal c-axis, and are not antiferroelectrically distorted. The local environment of each vanadium atom of the undoped tetragonal structure is identical to experimentally determined environment of the tungsten in doped VO$_{2}$. Thus this arrangement in the doped structure represents a step back towards the metallic tetragonal form. A resolution of this apparent paradox may be proposed by recognizing that the driving force for the symmetrization may be electron localization of the usual Mott-Hubbard type.
 
\begin{figure}[h!]
  \subfigure{\includegraphics[width=0.75\columnwidth]{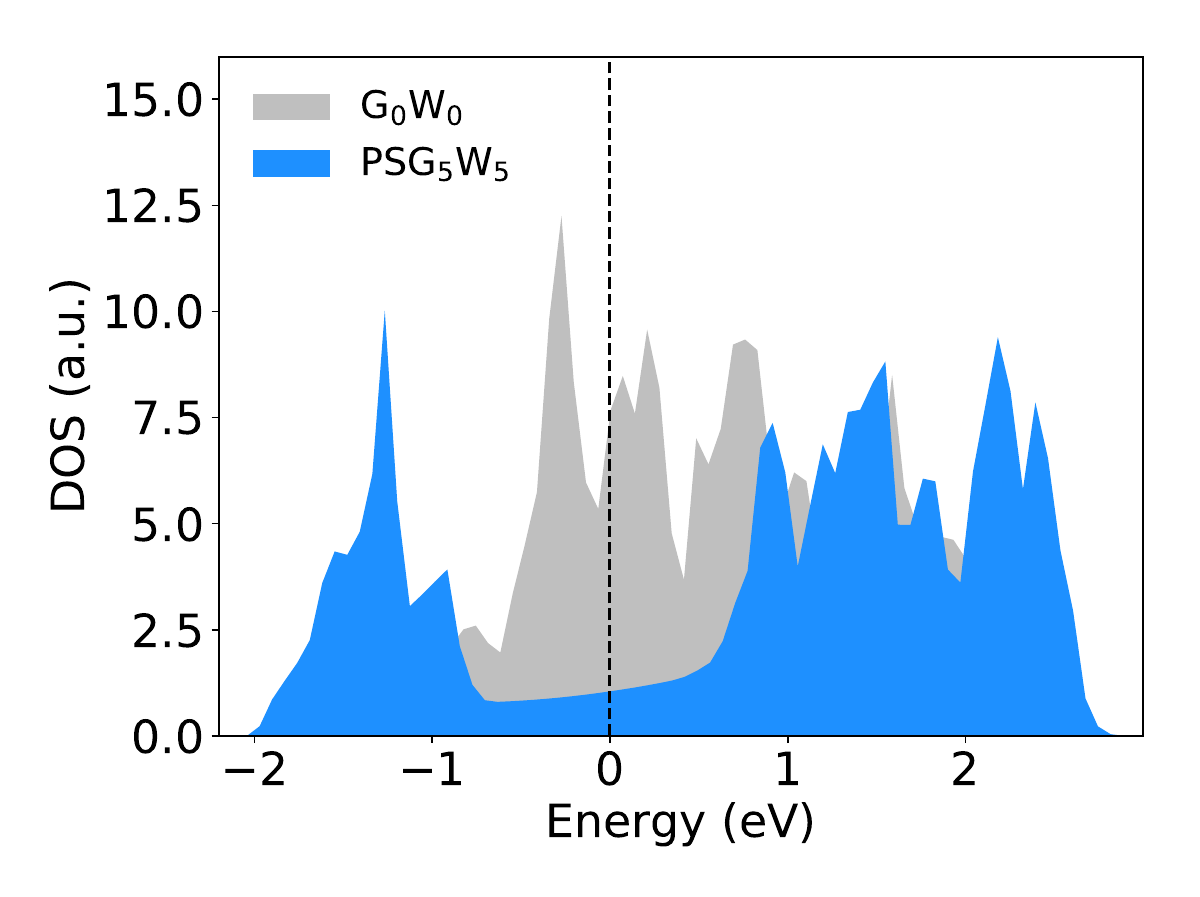}}\textbf{\textsf{a}}
  \subfigure{\includegraphics[width=0.9\columnwidth]{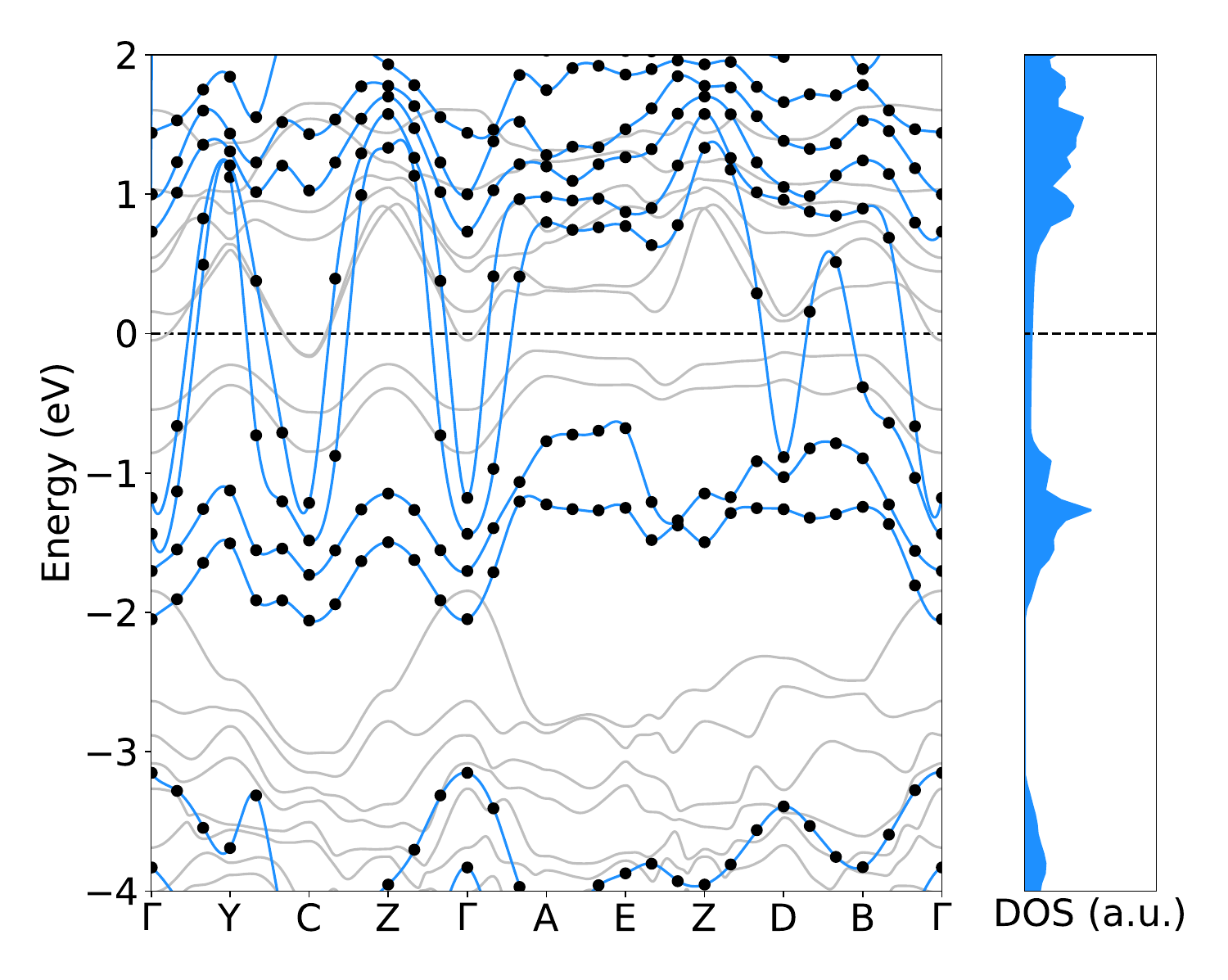}}\textbf{\textsf{b}}
  \subfigure{\includegraphics[width=0.75\columnwidth]{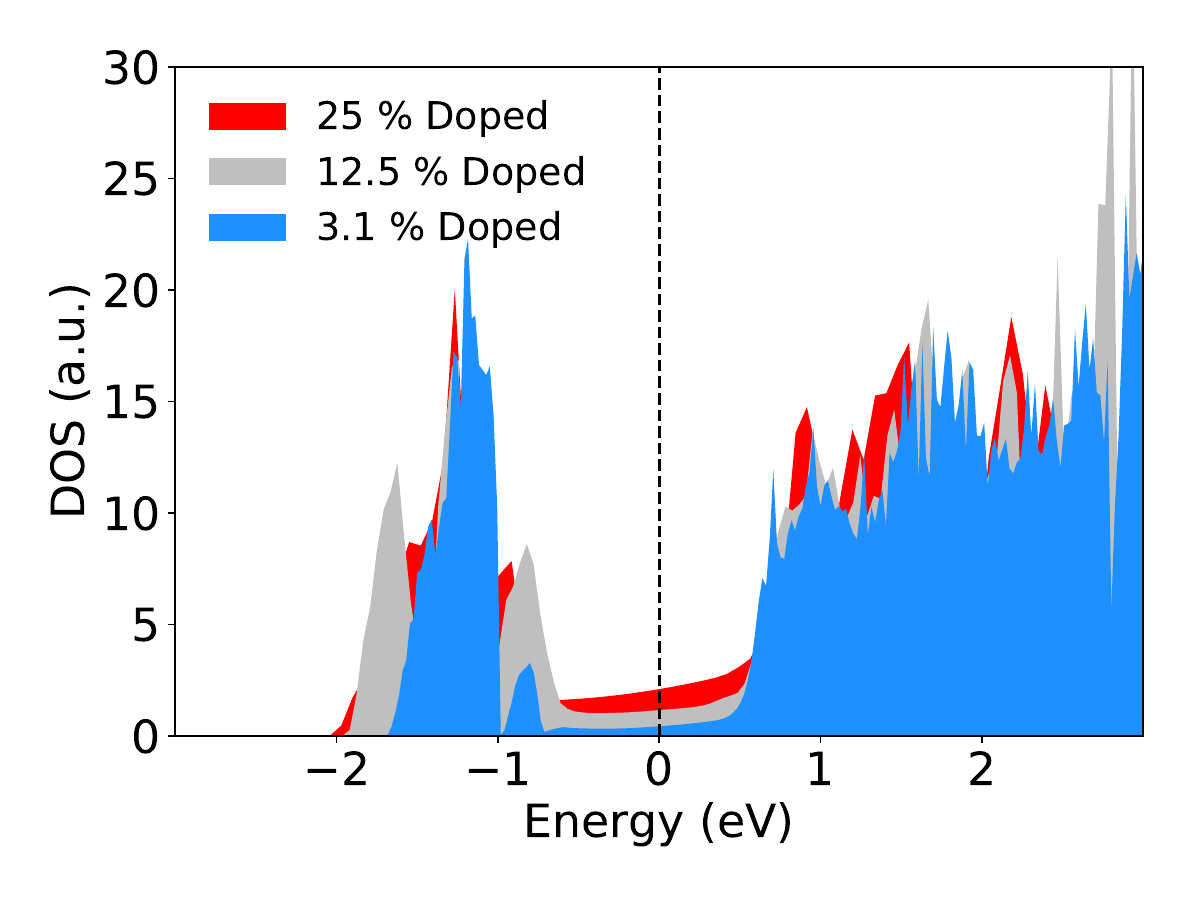}}\textbf{\textsf{c}}
  \caption{\raggedright{Color Online a) G$_{0}$W$_{0}$ (gray fill) and PS-G$_{5}$W$_{5}$ (blue fill) DOSs of tungsten doped VO$_{2}$ with internuclear symmetrisation around the dopant atom, b) Left panel: PS-G$_{5}$W$_{5}$ (black circles) and DFT (gray lines) band structures of the symmetrised, doped structure, Right panel: corresponding G$_{5}$W$_{5}$ DOS, and c) comparison of the DOSs of VO$_{2}$ with dopant percentages of 25 \% (red fill), 12.5 \% (gray fill) and 3.1 \% (blue filled curve).}}
\label{fig:Fig4}
\end{figure}

Figure \ref{fig:Fig4} investigates the effect of this by comparing standard G$_{0}$W$_{0}$ data with a more strongly correlated PS-G$_{5}$W$_{5}$ calculations. Figure \ref{fig:Fig4}a compares the densities of states of the standard GW calculation of the experimentally observed structure with a calculation in  which strong local \textbf{k}-space correlations are introduced. In comparison to the G$_{0}$W$_{0}$ data, the strongly correlated calculation reveals a clear splitting of the density of states into peaks equidistant from the Fermi level. The band structure of Figure \ref{fig:Fig2}b reveals that these strong correlations push the empty conduction states up above the Fermi level while the filled states are stabilized in the opposite direction, resulting in the formation of the Hubbard bands. As far as insulating structures go, while some states still exist at the Fermi level, this seems a far more promising candidate. 

This symmetrization of the local environment of the tungsten dopant observed in photoemission experiments \cite{Tang1985,Booth2009a} (see Figure \ref{fig:Fig1}c) can be reconciled by starting from a Mott insulating ansatz. From Figures \ref{fig:Fig2}b-c, the most significant effect of introducing a tungsten donor is to dope an electron into the structure, pushing the Fermi level up. From Figure \ref{fig:Fig2}a, it is clear that undoped M$_{1}$ VO$_{2}$ already has a filled valence band, therefore tungsten doping introduces an electron into the system, creating a partially filled band. If the tungsten atom Peierls pairs with a vanadium atom, the extra electron would interact with both electrons of the Peierls pairing, due to the decrease the hopping energy by reducing the internuclear spacing, and creating potential overlap between the atoms. However if the structure symmetrizes, the two valence electrons on the tungsten atom localize to it, and the single $d$-electron of the vanadium localizes on the vanadium site, reducing interactions. In this respect, this symmetrization is a classic Mott transition; increasing internuclear spacing giving rise to smaller correlations.

The W$_{0.25}$V$_{0.75}$O$_{2}$ doped structure of Figure 4a-c corresponds to a dopant percentage of 25 \%, which is much higher than those commonly used to adjust T$_{c}$. At this percentage the structure would be expected to be metallic experimentally \cite{Tang1985}. Figure 4c explores the effect of reducing the dopant concentration from 25 \% through 12.5 \% to 3.1 \% (scaled such that they integrate to the same total number of states). As the dopant percentage decreases, the density of states at E$_{F}$ decreases concurrently, suggesting that the structure becomes increasingly more insulating as the dopant amount tends to zero, consistent with the fact that fewer carriers are being doped into the structure. This data indicates that at dopant percentages designed to drop T$_{c}$ to approximately room temperature ($\sim$ 2-3 \% \cite{Lawton1995}), tungsten doped VO$_{2}$ will exhibit insulating character.

In summary, by including strong correlations in GW calculations on tungsten-doped VO$_{2}$ it becomes evident that doping carriers into the band insulating VO$_{2}$ structure results in a switch to Mott insulating behavior. The added electron occupies the conduction band minima, and strong local correlations in \textbf{k}-space split these minima from the rest of the bands, while a concurrent disruption of the Peierls pairing created by the dopant atom results in them combining with the original valence band into a broad lower Hubbard band. The data suggests that a modification of the pure VO$_{2}$ structure which results in an increase in electron number will create a Mott insulating defect band in which the Peierls pairing is disrupted, and the structure seeks to minimize correlations by symmetrizing the hopping distances around the defect. 

\textit{\textbf{Acknowledgements}}: SPR and JMB acknowledge the support of the ARC Centre of Excellence in Exciton Science (CE170100026). DWD acknowledges the support of the ARC Centre of Excellence for Nanoscale BioPhotonics (CE140100003). This work was supported by computational resources provided by the Australian Government through the National Computational Infrastructure National Facility and the Pawsey Supercomputer Centre.

Correspondence and requests for materials should be addressed to JMB, email: jamie.booth@rmit.edu.au

\bibliography{C:/Local_Disk/GWApproximation/Bibliography/library}

\end{document}